# Simultaneous Generation of Direct- and Indirect-Gap Photoluminescence in Multilayer MoS$_2$ Bubbles

Hailan Luo[1,2,♦], Xuanyi Li[1,2,♦], Yanchong Zhao[1,2,♦], Rong Yang[1], Yufeng Hao[3,4], Yunan Gao[5], Norman N. Shi[6], Yang Guo[1], Guodong Liu[1], Lin Zhao[1], Qingyan Wang[1], Zhongshan Zhang[1], Jiatao Sun[1*], Xingjiang Zhou[1,2,7,8*] and Yuan Huang[1,7*]

[1]*Beijing National Laboratory for Condensed Matter Physics, Institute of Physics, Chinese Academy of Sciences, Beijing 100190, China*

[2]*University of Chinese Academy of Sciences, Beijing 100049, China*

[3]*National Laboratory of Solid State Microstructures, College of Engineering and Applied Sciences, Jiangsu Key Laboratory of Artificial Functional Materials and Collaborative Innovation Center of Advanced Microstructures, Nanjing University, Nanjing 210093, China*

[4]*Haian Institute of New Technology, Nanjing University, Haian, 226600, China*

[5]*State Key Laboratory for Mesoscopic Physics, School of Physics, Peking University, Beijing 100871, China* [6]*Department of Applied Physics and Applied Mathematics, Columbia University, New York 10027, United States*

[7]*Songshan Lake Materials Laboratory, Dongguan 523808, China*

[8]*Beijing Academy of Quantum Information Sciences, Beijing 100193, China*

[♦]*These people contributed equally to the present work.*

* Corresponding authors: yhuang01@iphy.ac.cn, jtsun@iphy.ac.cn and xjzhou@iphy.ac.cn (Dated: January 19, 2020)

**Transition metal dichalcogenide (TMD) materials have received enormous attention due to their extraodinary optical and electrical properties, among which MoS$_2$ is the most typical one. As thickness increases from monolayer to multilayer, the photoluminescence (PL) of MoS$_2$ is gradually quenched due to the direct-to-indirect band gap transition. How to enhance PL response and decrease the layer dependence in multilayer MoS$_2$ is still a challenging task. In this work, we report, for the first time, simultaneous generation of three PL peaks at around 1.3, 1.4 and 1.8 eV on multilayer MoS$_2$ bubbles. The temperature dependent PL measurements indicate that the two peaks at 1.3 and 1.4 eV are phonon-assisted indirect-gap transitions while the peak at 1.8 eV is the direct-gap transition. Using first-principles calculations, the band structure evolution of multilayer MoS$_2$ under strain is studied, from which the origin of the three PL peaks of MoS$_2$ bubbles is further confirmed. Moreover, PL standing waves are observed in MoS$_2$ bubbles that creates Newton-Ring-like patterns. This work demonstrates that the bubble structure may provide new opportunities for engineering the electronic structure and optical properties of layered materials.**

Ultrathin two-dimensional (2D) materials exhibit fascinating electrical, optical and me- chanical performance. How to control and utilize their properties is important for various applications. MoS$_2$ as one of the most well studied 2D semiconductors shows high on-off ratio in field-effect transistor and strong photoluminence (PL), which has great potential in logic circuit, photodetector and flexible/wearable devices[1–5]. In the past decade, much attention has been focused on exploring the band structure and optical properties of mono- layer and bilayer MoS$_2$[6, 7]. Some multilayer related properties, like photoluminescence, are rarely reported. While the direct band gap nature of monolayer MoS$_2$ is important for its application in photoelectric devices, however, its layer dependence also brings challenges during material synthesis. If the unique properties in monolayer can be also realized in multilayer MoS$_2$, the processes of material synthesis and device fabrication will be greatly simplified. However, the goal for eliminating layer dependence is still a challenge in 2D ma- terials society. One strategy to prepare monolayer-like multilayer is to weaken the interlayer coupling by strain engineering[8, 9]. Bubble structure as recently used in strain engineering can provide a nonuniform strain field on layered materials[10–13], which delivers some inspi-ration for decoupling the interlayer interaction in multilayer MoS$_2$, as well as other multilayer TMD materials.

In this work, large size multilayer MoS$_2$ bubbles are fabricated on 300 nm SiO$_2$/Si sub- strate by a modified mechanical exfoliation technique[14]. Interestingly, while there is no PL signal on the flat region, three main PL peaks at 1.3 eV, 1.4 eV and 1.8 eV are clearly detected on multilayer MoS$_2$ bubbles. PL oscillations induced by optical standing waves are observed on multilayer MoS$_2$ bubbles for the first time. The origin of the three PL peaks is unambiguously identified by temperature-dependent PL measurements. The two lower energy PL peaks are phonon-assisted indirect gap transitions, which only emerge at higher temperature (>60 K). The PL peak at 1.8 eV still exists at low temperature (20 K), confirming it is a direct band transition. First-principles calculation results unveil the electronic structure changes of multilayer MoS$_2$ under biaxial strain, from which the origin of the three PL peaks can be also confirmed.

Figure 1a and 1b show the schematic and optical microscope images of a typical large MoS$_2$ bubble with lateral size of ~ 40 μm exfoliated on SiO$_2$/Si substrate. Newton ring patterns caused by white light interference are clearly seen in Fig. 1b. Large MoS$_2$ bubbles with diameters up to 60 μm can be obtained (more bubble images are presented in Fig. 2a), which are much larger than other MoS$_2$ bubbles transferred onto hole-array substrates[15–17]. Limited by the bubble size, some exotic phenomena such as Raman and PL oscillation reported here cannot be observed before[15–17]. Therefore, large size bubble is crucial for exploring various new properties of layered materials. The thickness of this flake has been identified by atomic force microscope (AFM) as shown in Fig. S1 (~ 8 nm), from which the number of MoS$_2$ layers can be determined (~ 13 layers). Tensile strain (up to 2.1%) in few-layer MoS$_2$ bubble samples have been confirmed by AFM measurement (Fig. S2). Compared with monolayer MoS$_2$ bubbles transferred on hole-array substrates, our multilayer MoS$_2$ bubbles exfoliated on flat SiO$_2$/Si substrate can sustain higher pressure difference, which helps to prepare big bubbles with larger strain. It is also worth pointing out that the shape of these MoS$_2$ bubbles can stay intact after one year. The outstanding stability of these MoS$_2$ bubbles shows a promising prospect for applications, such as micro-lens. Fig. 1c shows the PL spectra of multilayer MoS$_2$. Three PL peaks at 1.30, 1.45 and 1.74 eV are clearly detected on bubble area. As expected, no clear PL signal can be observed on flat multilayer MoS$_2$ area because of the indirect band gap nature in multilayer MoS$_2$.

The emergence of PL on multilayer MoS$_2$ bubble could be the result of interlayer decoupling effects, which means multilayer MoS$_2$ can exhibit some properties similar to monolayer MoS$_2$ once interlayer interaction is weakened. Raman spectroscopy is an useful tool to characterize the shearing and coupling between layers[18, 19]. In order to understand the interlayer behaviors in bubbles, we carried out Raman measurement for the bubble in Fig. 1b. The redshift of $E^1_{2g}$ indicates that the bubble sustains tensile strain (Fig. 1d). The full width at half maximum (FWHM) of the $A_{1g}$ peak at flat area is 2.6 $cm^{-1}$. However, the FWHM of this peak on the bubble increases to 3.5 $cm^{-1}$ (Fig. 1d). The FWHM change of $A_{1g}$ mode in Raman spectra indicates that the interlayer coupling becomes weaker on the bubble area. On the multilayer MoS$_2$ bubble, the deformation of neighboring layers are different, so there will be sliding between layers, resulting in change of the shearing mode in the multilayer. In the Raman spectrum of multilayer MoS$_2$, the $E^2_{2g}$ mode at the low-frequency range (<50 $cm^{-1}$) is sensitive to the shear vibration. Here, the intensity of the $E^2_{2g}$ mode on bubble is three times higher than that of the flat area, indicating that the shear vibration mode has been enhanced on the bubble region.

Figure 2a and 2b show the optical and PL mapping images of five MoS$_2$ bubbles with different diameters, which are integrated at 1.75 eV. From the PL intensity mapping images, we can clearly see the oscillation behavior on MoS$_2$ bubble gradually emerges as the size of bubble increases. The oscillation phenomenon originates from standing waves. When a MoS$_2$ bubble on Si substrate is irradiated by laser light, standing waves can be formed by the interference of incident and reflected beams, generating interference maxima at certain distances from the MoS$_2$-substrate interface. Similar oscillation behaviors were discovered in Raman mapping images of graphene bubbles[12]. However, there are still some obvious differences between graphene and MoS$_2$ bubbles. Graphene is a semimetal while MoS$_2$ is a semiconductor which can show more PL induced effects. To identify the origin of PL oscillation, PL mappings are displayed in Fig. 2b and Fig. S3. It can be clearly found that the PL ring's diameters and ring numbers are different in these images integrated at 1.3 eV, 1.4 eV and 1.7 eV. First, we extract 19 PL spectra along the yellow dashed line as indicated on the third bubble shown in Fig. 2b, which all show three PL peaks in each spectrum (Fig. 2c). The PL intensity oscillation and peak position (~ 1.75 eV) of the bubble can be extracted, as shown in Fig. 2d. More bubbles' PL mappings integrated at three energies are also summarized in Fig. S4 and the corresponding variation trend of PL intensity and maximum peak position profiles across these centers of the bubble in Fig. 2b are shown in Fig. S5. For this bubble, the diameter is about 45 μm, and its height is ~ 1.4 μm, as shown in the AFM image in the inset of Fig. 2e. Fig. 2e shows the calculated positions of interference maxima on the bubble, by using the luminescence with a wavelength of 708 nm as the wavelength ($\lambda$) of light generating interference patterns. We find that the calculated positions of the bright rings fit well with experimental results, which means that the luminescence (708 nm, 1.75 eV) emitting from the bubble and the other beam of luminescence reflected from the $SiO_2$/Si surface generate interference rings. More evidences can be referred to in Fig. S3, which show the PL ring positions around 867 nm (1.43 eV) and 932 nm (1.33 eV), respectively. Therefore, the PL oscillation in the image is attributed to the interference of PL itself instead of incident laser. From the PL image, we know that the height difference between two bright rings is $\lambda/2$. Therefore, smaller bubbles with height lower than the half-wavelength can not generate interference.

To further distinguish the nature of the direct and indirect gap transitions in the mul- tilayer $MoS_2$ bubbles, we performed temperature dependent PL measurements on $MoS_2$ bubbles from 20 K to 300 K (shown in Fig. 3 and Fig. S6). The PL peak at ~ 1.8 eV can be clearly observed on the $MoS_2$ bubble between 20 K and 300 K. The PL peak at ~ 1.3 eV becomes detectable only when temperature is higher than 60 K. The other PL peak at ~ 1.4 eV starts to emerge as temperature increases to 150 K. Both of the two peaks at ~ 1.3 eV and ~ 1.4 eV become more prominent as temperature increases. The different temperature-dependent behaviors of the three PL modes prove that the peak at 1.8 eV is direct gap transition, while the other two PL modes are phonon-assisted indirect transitions. This assignment is consistent with our calculation results shown below.

We observe an obvious red shift that is nearly linear for ~ 1.8 eV PL as temperature increases from 20 K to 300 K (Fig. 3a and 3b). This trend can be described by the Varshni equation[20, 21]:

$$E_g(T) = E_g(0) - \alpha \times T^2/(T + \beta^2) \qquad (1)$$

which shows the direct band-gap change as a function of temperature. In this equation, $E_g(T)$ is the energy gap, $E_g(0)$ is its value at 0 K, and $\alpha$ and $\beta$ are constants. In addition, all the three peaks show intensity oscillation as shown in Fig. 3b. As temperature increases from 20 K to 300 K, the bubble also gradually expands since the trapped gas pressure increases with temperature (shown in Fig. S7). As a result, the intensity of PL peaks show maxima and minima oscillation alternately as the height of the bubble increases when the sample warms up.

The above temperature-dependent PL measurements, for a given position, simultaneously change both the local temperature and strain. To isolate only the temperature effect, we performed laser power dependent PL measurements on the bubble. The laser is focused on the bubble with a spot size of ~ 500 nm. In this case, the local temperature increases with increasing power, while the strain change is negligible. Fig. 3d and 3e show PL peaks of one $MoS_2$ bubble under different laser powers. We find that the positions of the two indirect-gap PL peaks show a slight energy shift, but the direct-gap PL peak obviously moves to lower energy direction as the laser power increases (Fig. 3f). We also observe that the intensity of indirect-gap PL peaks, especially the one at 1.4 eV, increases linearly with increasing laser power (Fig. S8).

To further reveal the origin of the three PL transitions on multilayer $MoS_2$ bubbles, we performed first-principle band structure calculations. The electronic band structures of bilayer $MoS_2$ are calculated by varying the in-plane biaxial strain (left panel in Fig. 4a) and the interlayer distance (right panel in Fig. 4a), respectively. We found that three electronic excitations, namely direct transition at K point, indirect transitions at $\Gamma \rightarrow K$ and $\Gamma \rightarrow \Lambda$, are dominant around the Fermi level (Fig. 4a). Their band gaps decrease as the lattice biaxial strain increases with constant interlayer separation. On the other hand, the indirect excitation gaps increase significantly and the direct excitation gap increase just slightly as the interlayer separation increases (Fig. 4b). The contrasting behavior between the biaxial strain and interlayer separation also emerges in six-layer $MoS_2$ (Fig. S9). When the bilayer $MoS_2$ has equilibrium interlayer separation, the tensile strained bilayer $MoS_2$ with smaller interlayer separation has indirect gap excitation of 1.30 eV and 1.45 eV, which is in agreement with experimental PL data (Fig. 1c). We found that the direct band gap transition of 1.73 eV in bilayer $MoS_2$ at K point has little dependence on the interlayer separation even at larger lattice strain. Although our calculations are performed without considering electron-hole interaction and spin-orbit coupling, the strain dependence of band gap will not be affected, therefore, the calculation results can be well applied to explain the unusual PL behavior of multilayer $MoS_2$ bubbles.

In order to study the interlayer coupling directly, we calculated the charge density differences defined by $\Delta\rho=\rho_{bilayer}-\rho_{top}-\rho_{bottom}$, where $\rho_{bilayer}$, $\rho_{top}$, $\rho_{bottom}$ are the total charge density of the bilayer MoS$_2$, the separate charge density of top layer and bottom layer charge density, respectively. Here, we choose a bilayer MoS$_2$ with 3.8 Å interlayer distance for clear statement (Fig. 4c). Fig. 4d-f show the charge density difference between layers under a smaller (3.4 Å) and larger (4.0 Å) interlayer distance, tensile (+2%) and compressive (-2%) strain, and layer shift in armchair and zig-zag (0.5 Å) directions. The interlayer charge density becomes depleted when applying tensile strain or increasing the interlayer distance. On multilayer MoS$_2$ bubble, a tensile strain is applied on MoS$_2$ lattice, and the interlayer distance can also be increased. As shown above, the tensile strain and increased interlayer seperation can weaken the interlayer coupling in multilayer MoS$_2$ bubble, and as a result, some monolayer behaviors become observable on the multilayer bubble structure.

In summary, we observed both direct- and indirect-gap photoluminescence emissions in multilayer MoS$_2$ bubbles. AFM measurement demonstrated that the crystal lattice constant has been stretched on bubble, leading to weakened interlayer coupling. Three PL peaks have been discovered on multilayer MoS$_2$ bubbles due to the interlayer decoupling. Larger MoS$_2$ bubbles (diameter >5 μm) show PL oscillation in the mapping images owning to the inter- ference effect of PL. From temperature-dependent PL measurements, we verified that the two peaks at around 1.3 and 1.4 eV are phonon-assisted indirect-gap transitions. Moreover, the direct-gap PL red shifts as the temperature rises up. The DFT calculation results reveal the evolution of the band structure of multilayer MoS$_2$ with strain and interlayer separation, whick clarfied the origination of the three PL peaks on bubble. Multilayer MoS$_2$ bubbles provide new platform for understanding of the band structure evolution and optical property changes of this material under strain. This work can delivery new inspiration for the field of light-emitting diodes and photodetectors based on multilayer TMDCs.

**Acknowledgement**

This work is supported by the National Key Research and Development Program of China (Grant No. 2019YFA0308000, 2018YFA0305800, 2016YFA0300300), the National Natural Science Foundation of China (Grant No. 11888101, 11922414, 11874405, 11974045, 51772145), and the Strategic Priority Research Program (B) of the Chinese Academy of Sci- ences (Grant No. XDB25000000), the Research Program of Beijing Academy of Quantum Information Sciences (Grant No. Y18G06).

**Author Contributions**

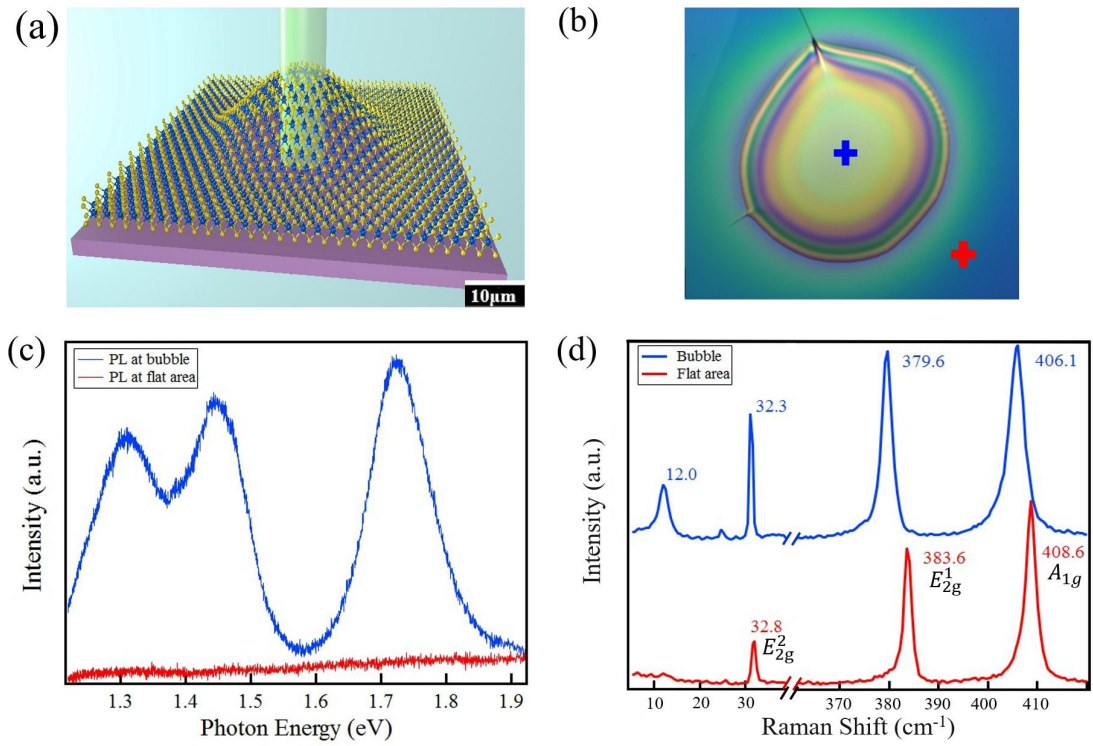

FIG. 1: **Photoluminescence response on multilayer MoS$_2$ bubble.** (a) Schematic diagram of a MoS$_2$ bubble exfoliated onto a Si wafer with a 532 nm laser beam irradiating on it. (b) Optical microscopy image of a multilayer MoS$_2$ bubble on a SiO$_2$/Si substrate showing Newton rings. (c) Photoluminescence spectra measured on MoS$_2$ bubble center (red curve) and the flat region (blue curve), respectively. The measuring positions are marked by blue and red crosses in Fig. 1b.

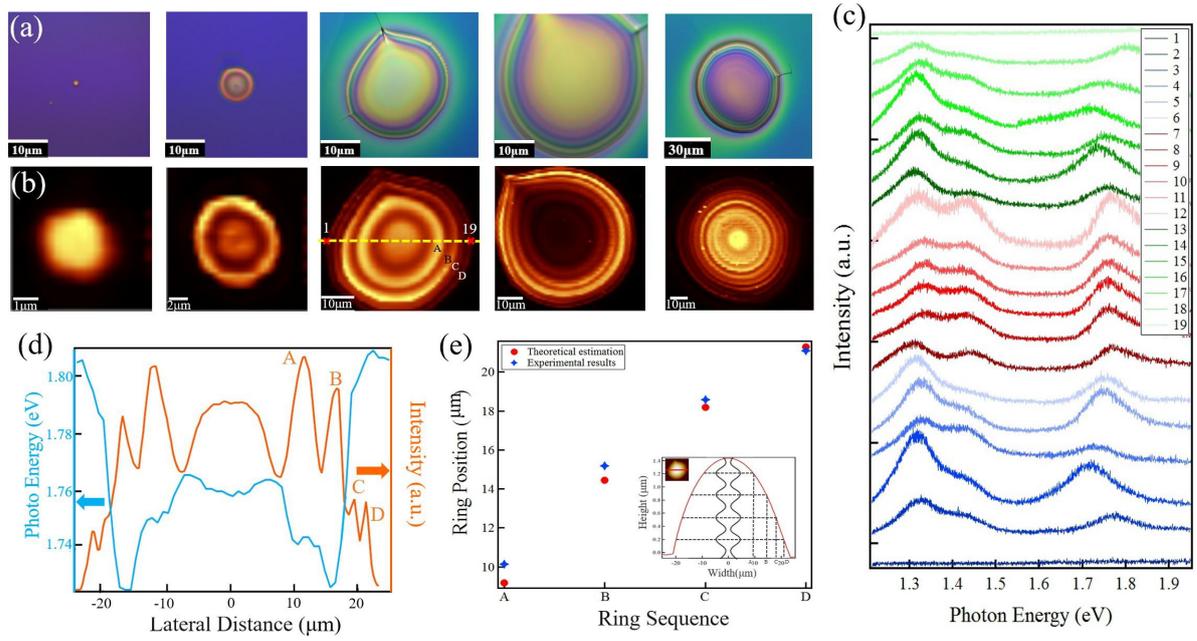

FIG. 2: **The PL oscillation behavior on multilayer MoS$_2$ bubble.** (a) Optical images and PL mapping images of five different multilayer MoS$_2$ bubbles. The five PL intensity mapping images are integrated at 1.75 eV (integration energy window: 0.1 eV). (b) PL spectra extracted across the third bubble's center, which are collected along the yellow dashed line shown in (a). The inset in (c): Height profile of the third bubble in (a), which is measured along the yellow dashed line. The positions on these circles with interference maximum amplitude are marked as A, B, C and D. (d) The oscillation of PL intensity and PL peak position of the third MoS$_2$ bubble in (a), which is extracted from (b).

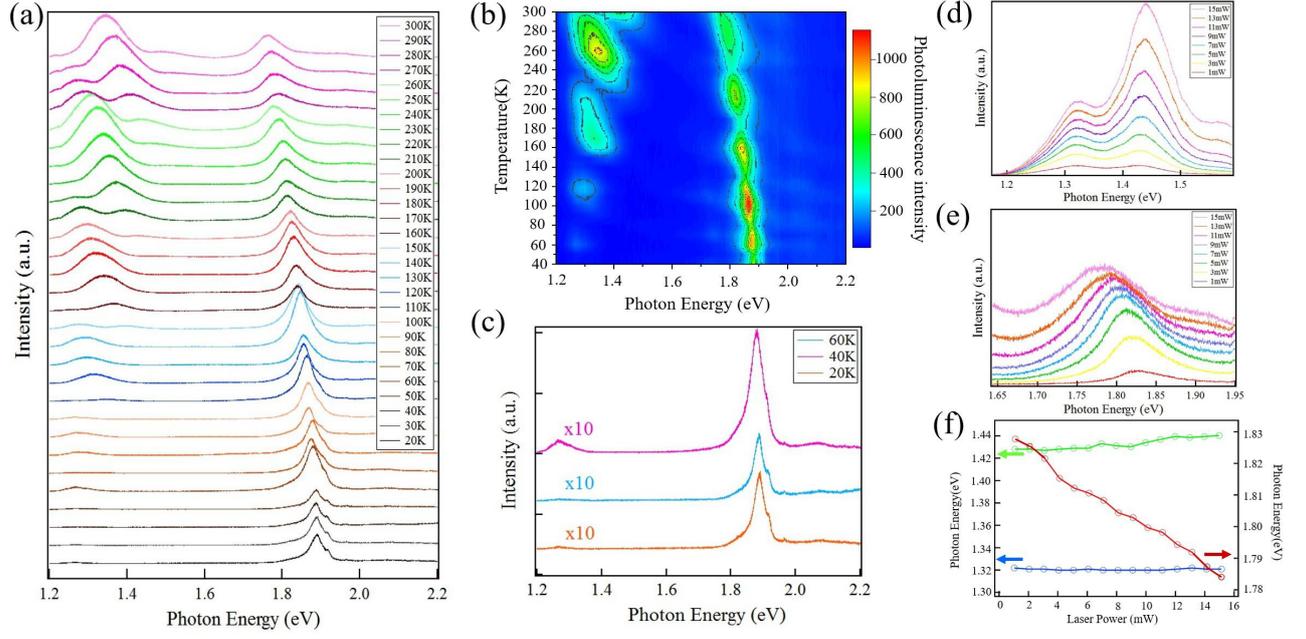

FIG. 3: **Temperature dependence of PL on MoS$_2$ bubble.** (a) PL spectra of a MoS$_2$ bubble in the temperature range from 20 K to 300 K. (b) 2D mapping image of PL intensity in (a), from which we can observe the direct PL at ~ 1.8 eV shows a red shift. (c) Three PL specra of MoS$_2$ bubble measured at 20 K, 40 K and 60 K. The indirect-gap PL peak around 1.3 eV become visiable as temperature increase to 60 K. (d and e) Laser power dependent of the indirect- and direct-gap PL spectra of another multilayer MoS$_2$ bubble. Both PL peaks gradually increase as laser power increase from 1 mW to 15 mW. (d) Peak position variation with the increasing laser power for the two indirect-gap PL peaks at ~ 1.3 eV, ~ 1.4eV and one direct-gap PL peak at ~ 1.8eV, extracted from (d and e).

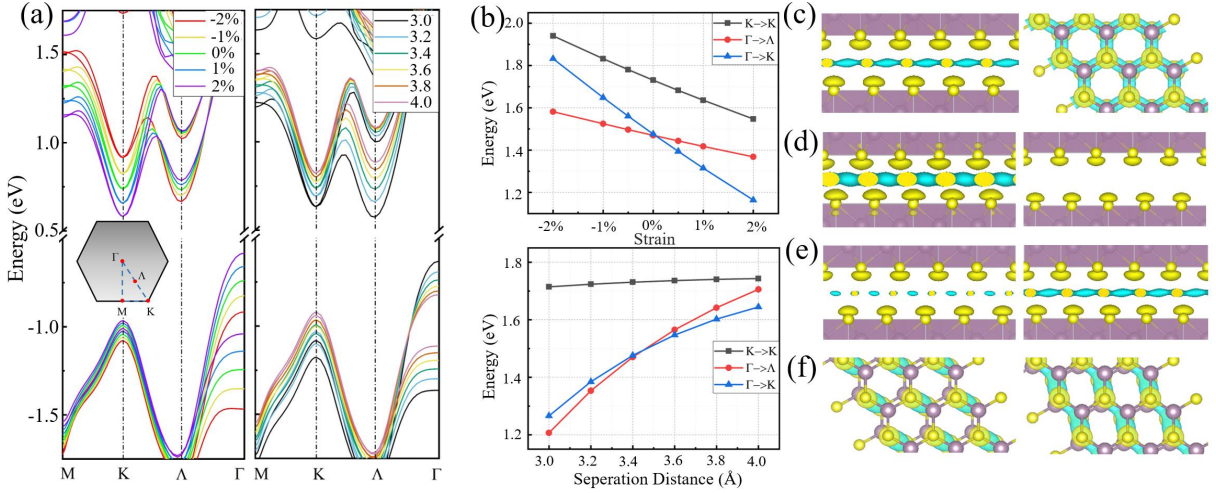

FIG. 4: **Band structure calculations on the origin of PL peaks on MoS$_2$ bubble.** (a) Band structures of bilayer MoS$_2$ with different strain (left) and interlayer distance (right). (b) Statistical diagram of band gap dependence of bilayer MoS$_2$ as a function of strain (upper) and interlayer distance (lower). The raw data are from panel (a). The charge density difference of bilayer MoS$_2$ that indicates the interlayer coupling are presented in c-f. (c) Strain-free bilayer MoS$_2$ with 3.8 Å separation of sulfur atoms in neighboring layers distance is taken as a reference system. (d) Side view of charge density difference by changing the interlayer distance (left: 3.4 Å, right: 4.0 Å). (e) Side view of bilayer MoS$_2$ under biaxial tensile (left: +2%) and compressive (right: -2%) strain. (f) Interlayers shift of 0.5 Å along armchair (left) and zigzag (right). All the plots of charge density difference have a same iso-surface value of $1.45 \times 10^{-4}$ e/Å

*Supplementary Materials for*

# Simultaneous Generation of Direct- and Indirect-Gap Photoluminescence in Multilayer MoS$_2$ Bubbles

**Hailan Luo[1,2]♦, Xuanyi Li[1,2]♦, Yanchong Zhao[1,2]♦, Rong Yang[1], Yufeng Hao[3,4], Yu-nan Gao[5], Norman N. Shi[6], Yang Guo[1], Guodong Liu[1], Lin Zhao[1], Qingyan Wang[1], Zhongshan Zhang[1], Jiatao Sun[1]\*, Xingjiang Zhou[1,2,7,8]\* and Yuan Huang[1,7]\***

[1]*Beijing National Laboratory for Condensed Matter Physics, Institute of Physics, Chinese Academy of Sciences, Beijing 100190, China*

[2]*University of Chinese Academy of Sciences, Beijing 100049, China*

[3]*National Laboratory of Solid State Microstructures,*

*College of Engineering and Applied Sciences,*

*Jiangsu Key Laboratory of Artificial Functional Materials and*

*Collaborative Innovation Center of Advanced Microstructures, Nanjing*

*University, Nanjing 210093, China*

[4]*Haian Institute of New Technology,*

*Nanjing University, Haian, 226600, China*

[5]*State Key Laboratory for Mesoscopic Physics, School of*

*Physics, Peking University, Beijing 100871, China* [6]*Department of*

*Applied Physics and Applied Mathematics,*

*Columbia University, New York 10027, United States*

[7]*Songshan Lake Materials Laboratory, Dongguan 523808, China*

[8]*Beijing Academy of Quantum Information Sciences, Beijing 100193, China*

♦*These people contributed equally to the present work.*

\* *Corresponding authors: yhuang01@iphy.ac.cn,*

*jtsun@iphy.ac.cn and xjzhou@iphy.ac.cn* (Dated:

January 19, 2020)

**Methods**

**Experiments. Fabrication of MoS$_2$ bubbles.** Before exfoliating thin MoS$_2$, the SiO$_2$/Si substrate was exposed to oxygen plasma to remove adsorbates from its surface. Following this step, the MoS$_2$-loaded tape adhered to the substrate and the substrate was heated for 1~ 2 min at ~ 110 °C in air. Since the substrate surface can absorb small molecules in air, the interface between MoS$_2$ and substrate traps some of these gas molecules. These molecules can then accumulate at the interface and form bubble structures when being heated on a hot plate. After the sample cooled to room temperature, the adhesive tape was removed and bubbles with different sizes were fabricated[1].

AFM, Raman and PL measurements: Atomic force microscopy (Park, XE7) and confocal Raman spectroscopy/microscopy (WITec alpha 300) were used to measure the properties and thickness of exfoliated MoS$_2$. A laser wavelength of 532 nm and spot size of ~ 0.5 μm was used to obtain Raman/PL spectra and spatially resolved Raman/PL mapping images. In the temperature dependent PL measurement, the MoS$_2$ bubble sample was loaded in a home-made vacuum chamber with temperature control system. The temperature can be continuously controlled from 20 K to room temperature.

**Calculations.** Density functional theory calculations. All first-principle calculations are performed on the base of density functional theory (DFT[2, 3]) via Vienna ab initio simulation package (VASP)[4]. We used the local density approximation (LDA[5]) form for exchange-correction functional to calculate the charge distribution and electronic structure of bilayer MoS$_2$. Ion cores are described by using the projector-augmented wave (PAW[6]) method. A 9×9×1 k-point grid generated by Monkhost-Pack is used. The energy cutoff is set to be 500 eV. The spin-orbit coupling (SOC) has been included as well.

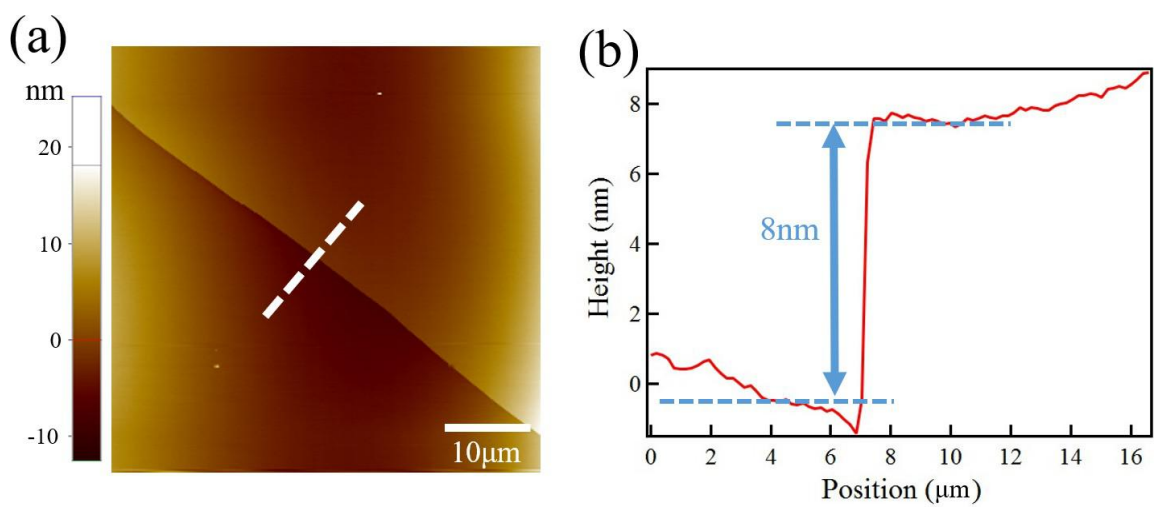

Fig. S1: (a) AFM image of MoS$_2$ plate edge. (b) Height profile of plate edge along the white dashed line in (a).

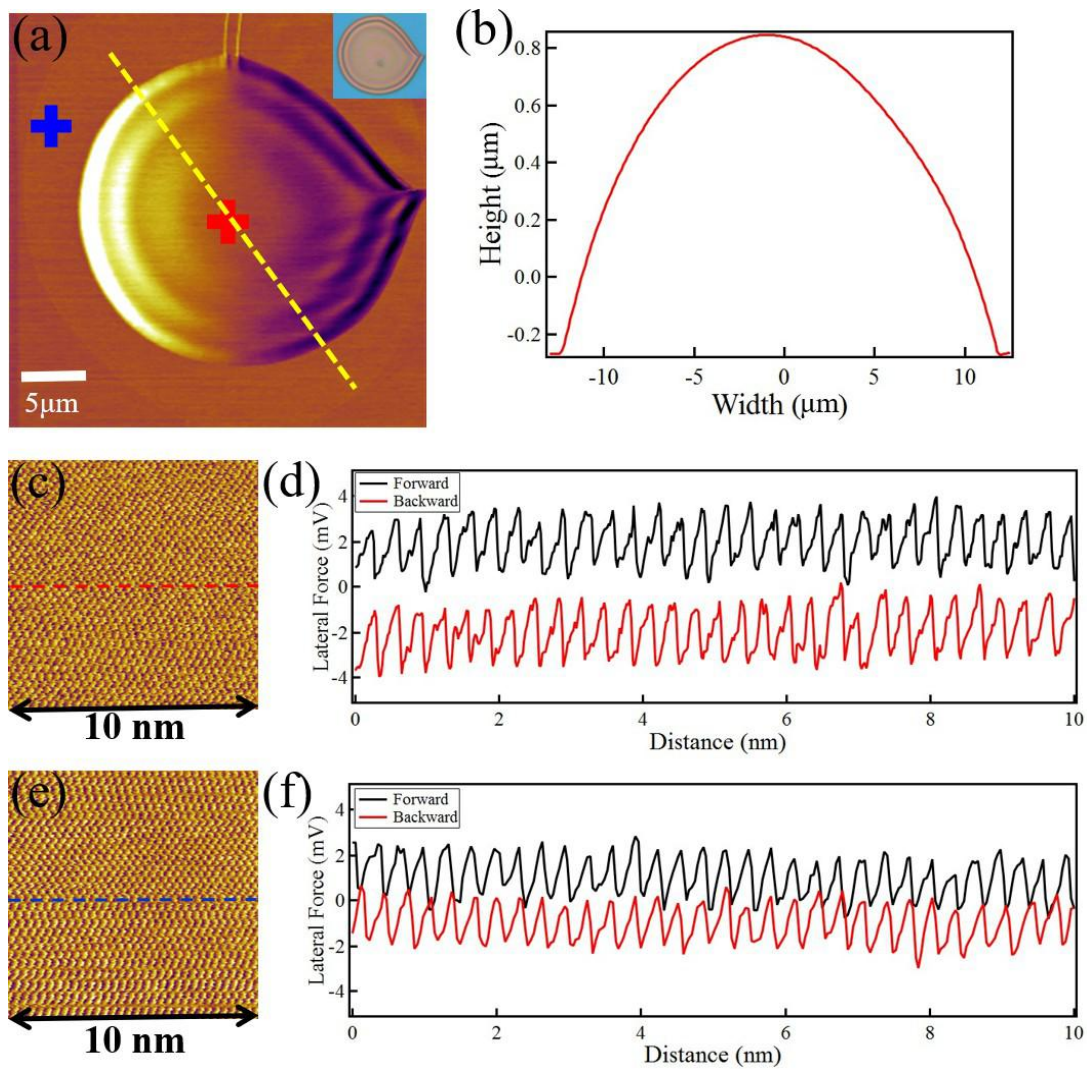

Fig. S2: (a) AFM image of one multilayer MoS$_2$ bubble. The corresponding optical microscope image is presented in the inset. (b) The height profile of the bubble which was measured across the center as shown in the dashed yellow line in (a). (c and e) Atomic resolution AFM image on the top of the bubble and on the adjacent flat area, which are marked by the red and blue cross, respectively. (d and f) The forward and backward line profile across the dashed line as shown in (c) and (e). The distance for 30 peaks in (d) is 9.72 nm, while the value is 9.52 nm for the curve in (f), from which the strain can be calculated to be 2.1%.

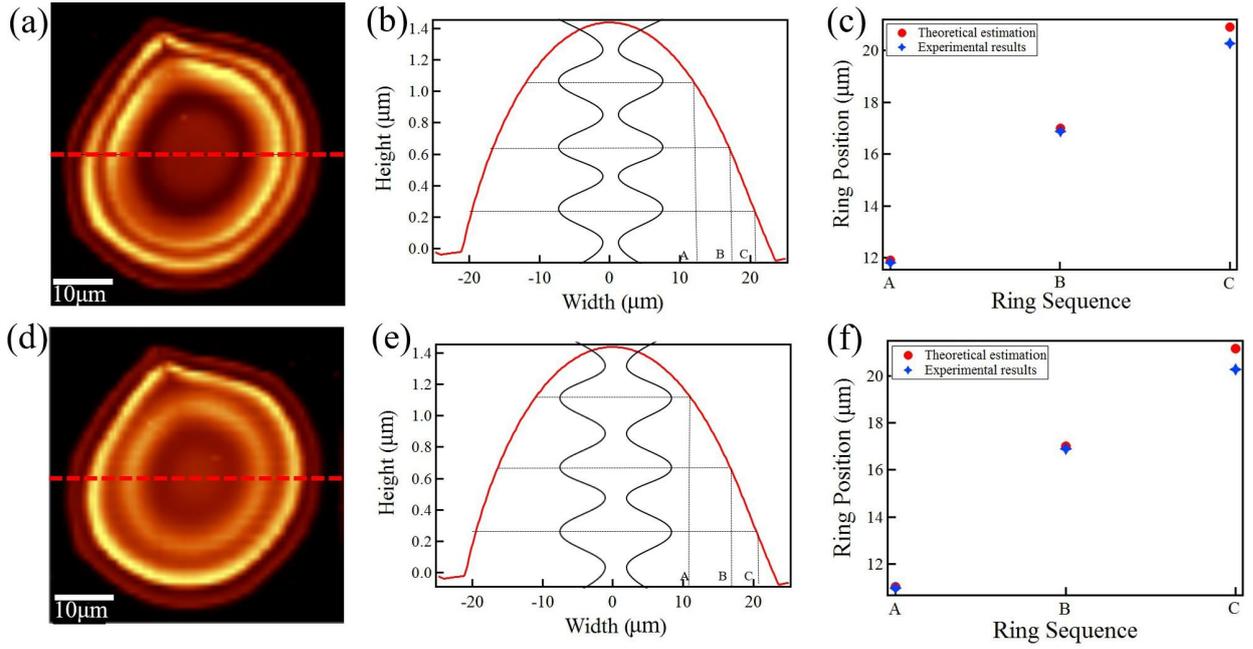

Fig. S3: (a and d) PL intensity image of the third bubble shown in Fig. 2a, which is integrated over an energy window of 0.1 eV, centered around 1.43 eV and 1.33 eV, respectively. (b and e) Schematic of a standing wave showing PL intensity maxima. The positions on these bright rings are marked by A, B, and C. (c and f) The comparison of the theoretical estimation and experimental measurements on the position of the bright rings.

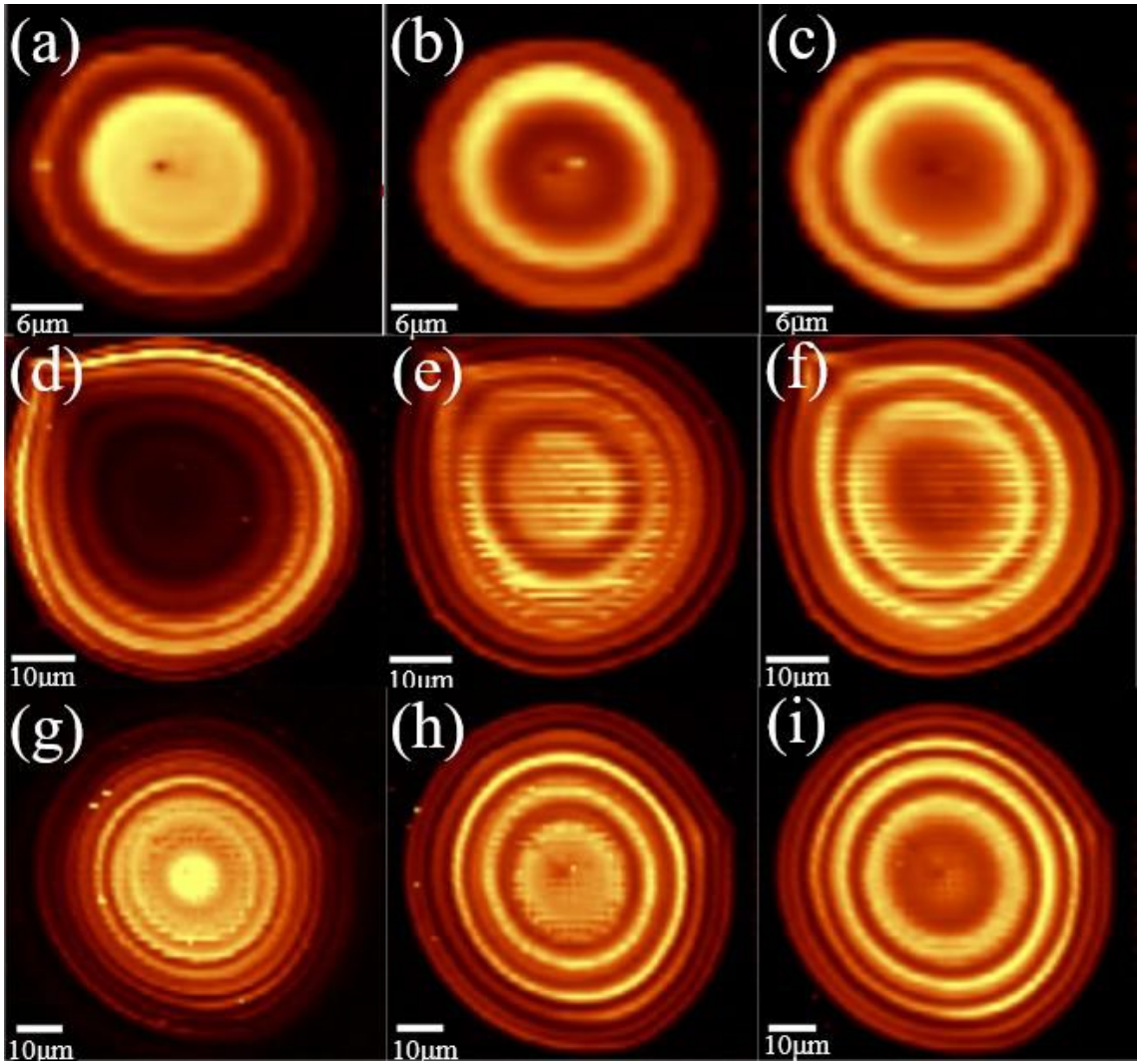

Fig. S4: (a)-(i) PL intensity mapping images of three bigger bubbles (shown in Fig. 2a), which are integrated over an energy range of 0.1eV, centered around 1.75 eV, 1.43 eV and 1.33 eV, respectively.

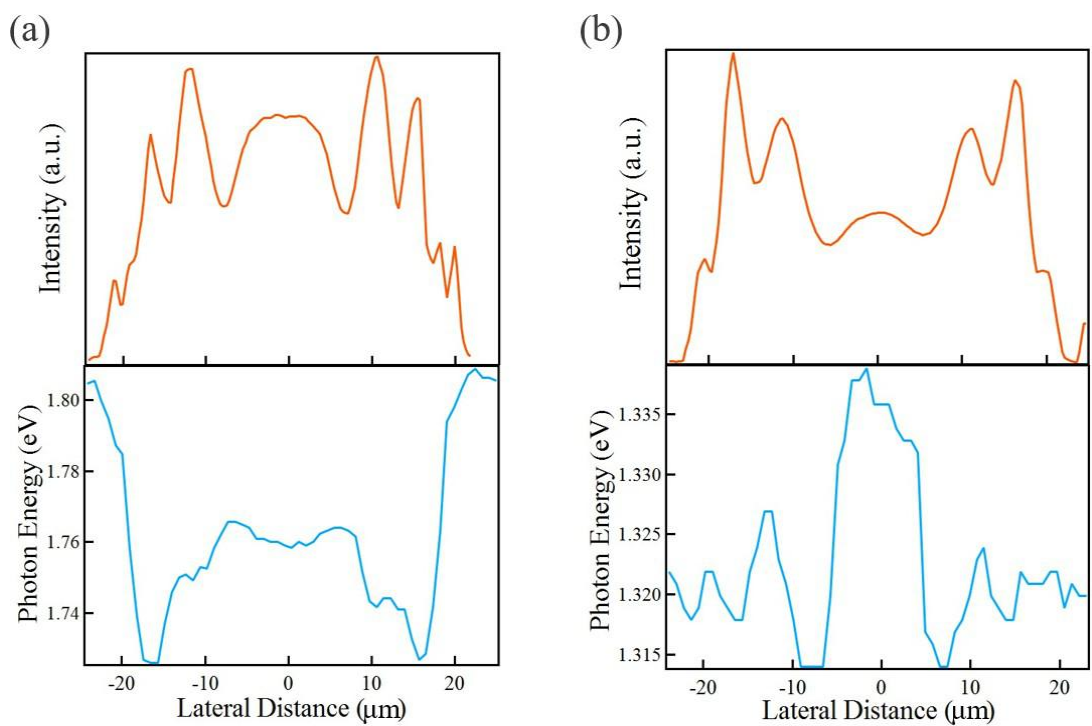

Fig. S5: The oscillation of PL intensity (top) and energy position (bottom) around 1.75 eV (a) and 1.33 eV (b) for the third $MoS_2$ bubble shown in Fig. 2a.

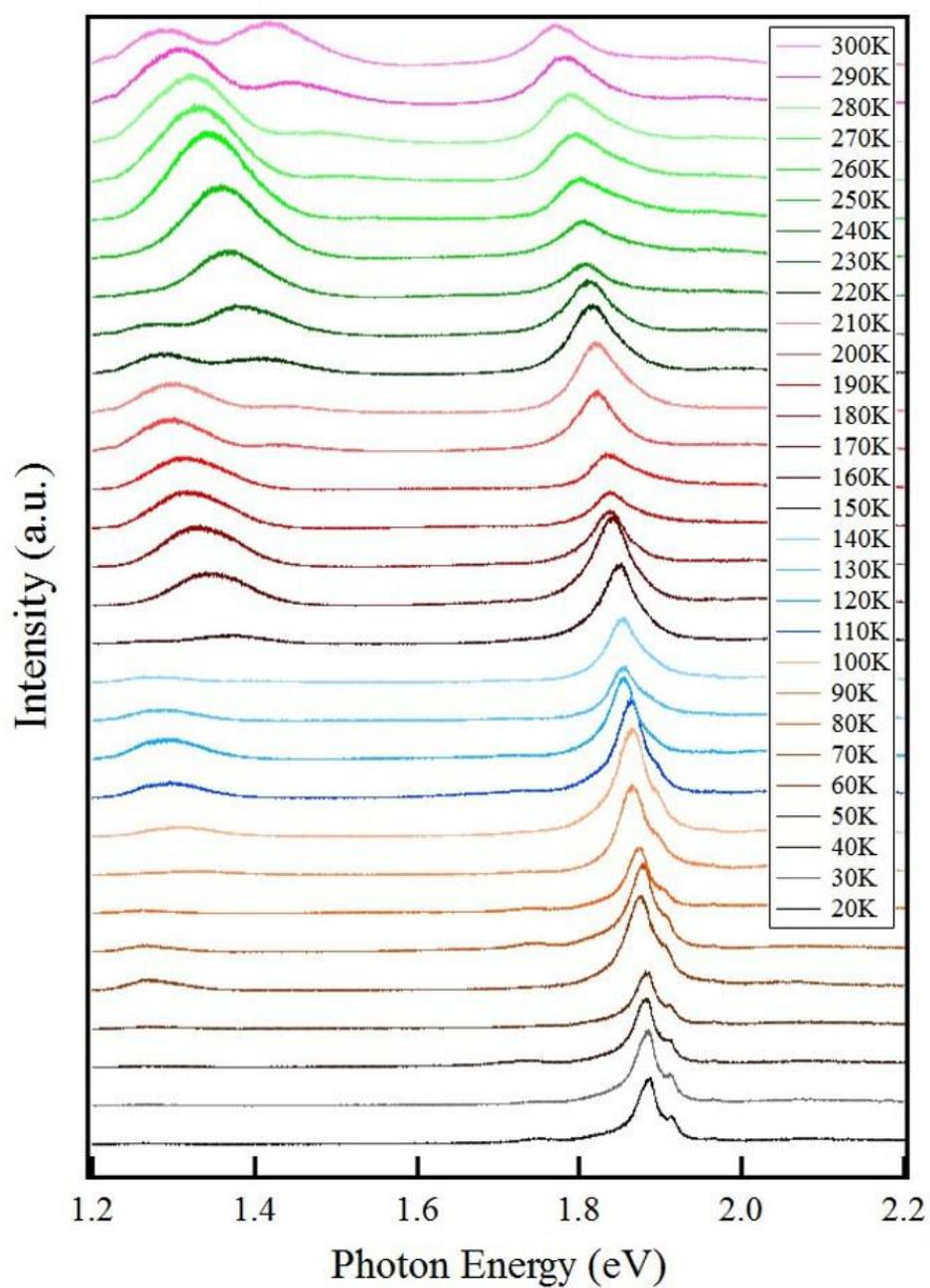

Fig. S6: The PL spectra of another MoS$_2$ bubble measured in the temperature range from 20 K to 300 K.

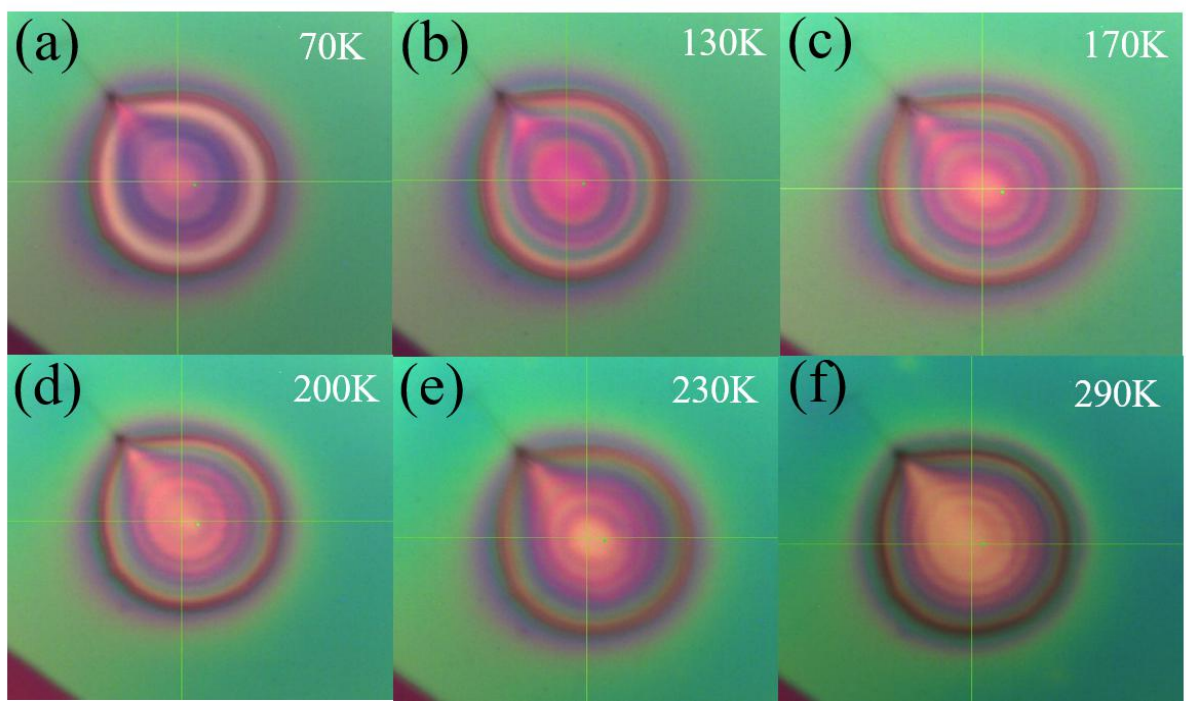

Fig. S7: (a-f) are the optical images of one $MoS_2$ bubble at different temperature, from which one can see the Newton's rings gradually change as temperature increase.

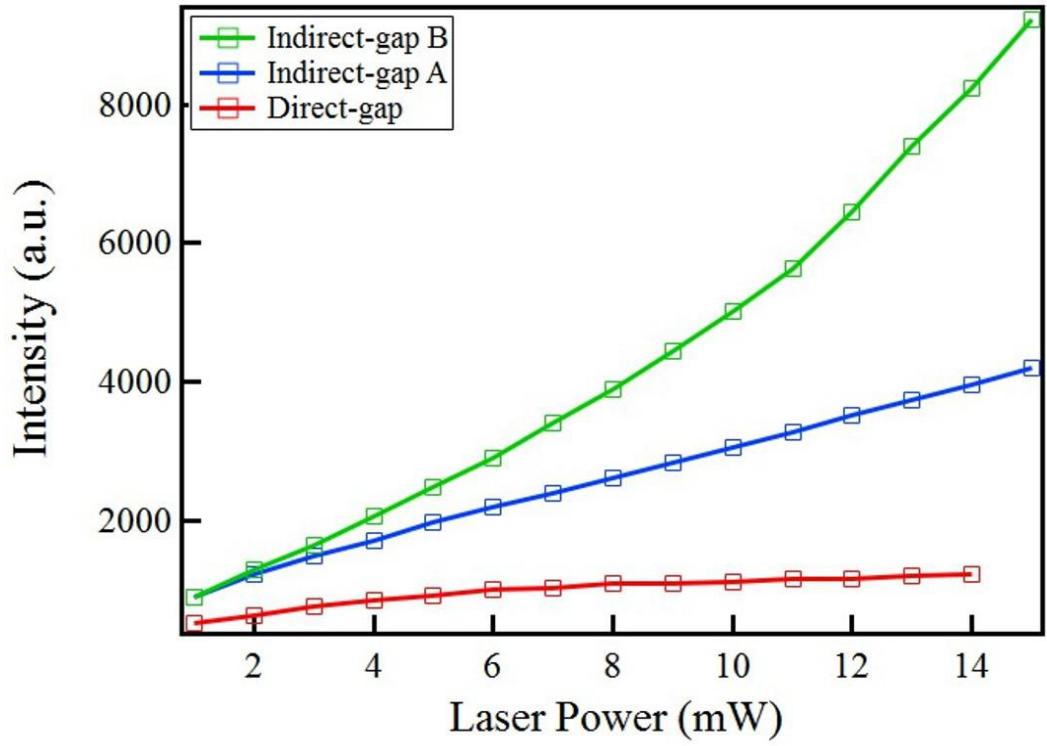

Fig. S8: The intensity variation of the two indirect-gap and one direct-gap PL peaks at different laser power, which is extracted from Fig. 3(d and e).

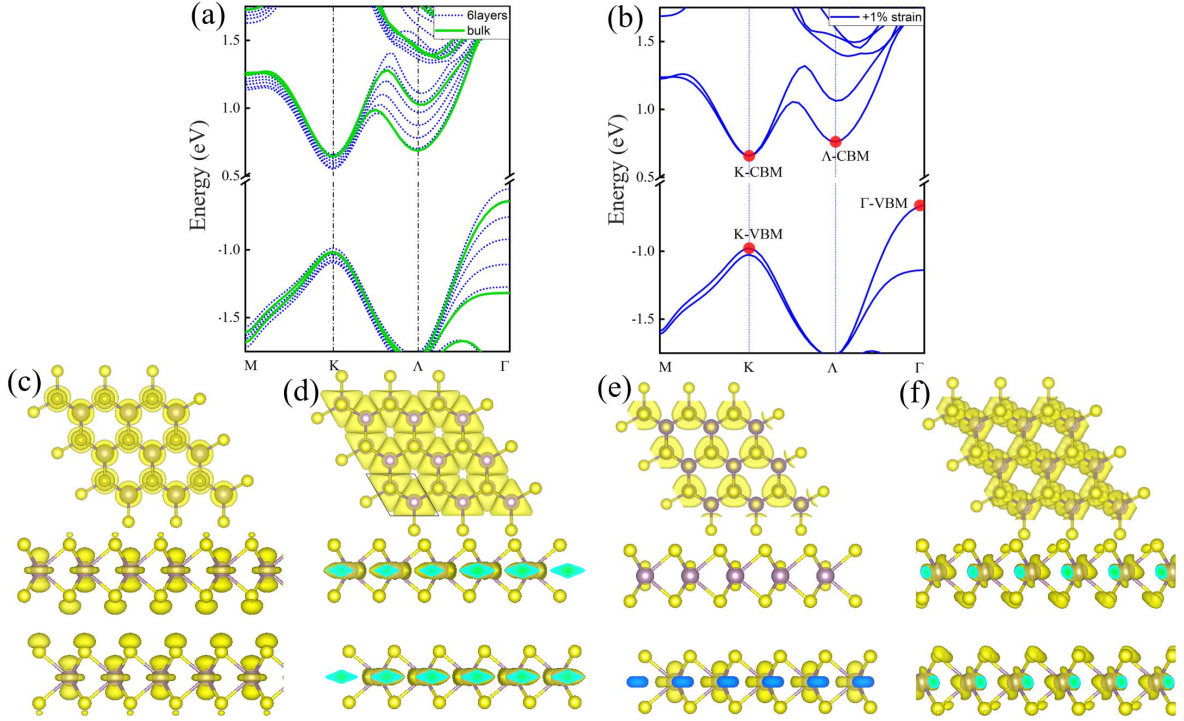

Fig. S9: (a) Band structure of 6-layer MoS$_2$ system, which shows similar shape when compared with the bulk. (b) Band structure of bilayer MoS$_2$ system under +1% tensile strain, some key points are marked by red dots. We modeled the charge distribution at these points as illustrated in figures (c-f). Top view and side view of the charge distribution of the partial charger at: (c) Γ point on VBM (d) K point on VBM (e) K point on CBM and (f) Λ point on CBM, under +1% tensile strain in a bilayer MoS$_2$ system.